# Novel Architecture of Pipeline Radix $2^2$ SDF FFT Based on Digit-Slicing Technique

[1]Yazan Samir Algnabi, [2]Furat A. Aldaamee, [3]Rozita Teymourzadeh,
[1]Masuri Othman, [1]Md Shabiul Islam

[1]Institute of Microengineering & Nanoelectronics (IMEN)
Universiti Kebangsaan Malaysia (UKM),
43600 UKM Bangi,
Selangor, Malaysia

[2]SCHOOL OF COMPUTER SCIENCE & IT,
Linton University College,
Bandar Universiti Teknologi Legenda
71700, Mantin,
N. Sembilan, Malaysia

[3]Faculty of Engineering, Technology and Built Environment,
Electrical & Electronic Engineering department, UCSI University, 56000,
Kuala Lumpur, Malaysia.

*Abstract* -- **The prevalent need for very high speed digital signals processing in wireless communications has driven the communications system to high performance levels. The objective of this paper is to propose a novel structure for efficient implementation for the Fast Fourier Transform (FFT) processor to meet the requirement for high speed wireless communication system standards. Based on the algorithm, architecture analysis, the design of pipeline Radix $2^2$ SDF FFT processor based on digit-slicing Multiplier-Less is proposed.** Furthermore, **this paper proposed an optimal constant multiplication arithmetic design to multiply a fixed point input selectively by one of the several present twiddle factor constants. The proposed architecture was simulated using MATLAB software and the Field Programmable Gate Array (FPGA) Virtex 4 was targeted to synthesis the proposed architecture. The design was tested in real hardware of TLA5201 logic analyzer and the ISE synthesis report results the high speed of** 669.277 **MHz with the total equivalent gate count of 14,854. Meanwhile, It can be found as significant improvement over Radix $2^2$ DIF SDF FFT processor and can be concluded that the proposed pipeline Radix $2^2$ DIF SDF FFT processor based on digit-slicing multiplier-less is an enable in solving problems that affect the most high speed wireless communication systems capability in FFT and possesses huge potentials for future related works and research areas.**

## I. INTRODUCTION

FFT plays an important role in many digital signals processing (DSP) application such as communication systems and image processing. It is an efficient algorithm to compute the discrete Fourier transform (DFT) and it's inverse. The DFT is main and important procedure in the data analysis, system design and implementation [1]. The challenge in FFT hardware implementation is the speed functionality of the multiplier unit. Hence, to reduce the complexity of the FFT calculation, many modules were developed [2-11]. However, in order to implement FFT processor as system on chip (SOC) ASIC implementation and FPGA prototyping were considered. Recently, FPGA has become an applicable option to direct hardware solution performance in the real time application. However, this paper will concentrate on FPGA implementation of high multiplier-less FFT processor using shift add technique. Since multiplication causes high delay propagation in FFT calculation, the new technique of digital-slicing is applied to build novel architecture of multiplier-less FFT processor. The motivation of this research work was inspired by [12-14, 18]. Meanwhile, the study of the digit-slicing FFT has been introduced by [15] in DSP application. Hence, this research will use a similar digit-slicing technique with the ones put forth by [15] but having a difference by the use of a different algorithm, architecture and different platform, which helps to improve the performance and achieve higher speed and performance.

## II. DIGIT SLICING ARCHITECTURE

The concept behind the digit-slicing is any complex number, *F*, can be sliced into smaller blocks, each having a shorter word length, *p*, as shown in the following equations [14].

$$F = \sum_{k=0}^{b-1}(2^{p-1})^k FR_k + j\sum_{k=0}^{b-1}(2^{p-1})^k FI_k \quad (1)$$

$$FR_k = -(2^{p-1})FR_{(p-1),i} + \sum_{i=0}^{p-2}(2^i)FR_{k,i} \quad (2)$$

$$FI_k = -(2^{p-1})FI_{(p-1),i} + \sum_{i=0}^{p-2}(2^i)FI_{k,i} \quad (3)$$

Where $FI_{k,i}$ and $FR_{k,I}$ have values which are either zero or one. Any value whose absolute value is less than one can be represented in two's complement as:

$$x = \left[\sum_{k=0}^{b-1}2^{pk}X_k\right]2^{-(pb-1)} \quad (4)$$

Where *x* is any number which its absolute value is less than one and *x* is sliced into b blocks of each p bits wide.

$$X_k = \sum_{j=0}^{p-1}2^j X_{k,j} \quad (5)$$

Where $X_{k,j}$ are all either ones or zeros except for $X_{k=b-1,\ j=p-1}$ which is either zero or minus one.

## III. RADIX $2^2$ SDF FFT ALGORITHM

The Radix $2^2$ FFT algorithm has the same multiplicative complexity as Radix 4 but retains the butterfly structure of Radix 2 algorithm [16]. In this algorithm, the first two steps of the decomposition of Radix 2 DIT-FFT are analyzed, and common factor algorithm is used to illustrate.

$$X[k] = \sum_{n=0}^{N-1} x[n] W_N^{nk}, k = 0,1,\dots,N-1 \quad (6)$$

In Eq. 6 the index n and k decomposed as:

$$n = <\frac{N}{2}n_1 + \frac{N}{4}n_2 + n_3 > N \quad (7)$$
$$k = <k_1 + 2k_2 + 4k_3 > N \quad (8)$$

The total value of *n* and *k* is *N*. When the above substations are applied to (6) the DFT definition can be written as the:

$$X[k_1 + 2k_2 + 4k_3] = \sum_{n_3=0}^{(N/4)-1} \sum_{n_2=0}^{1} \sum_{n_1=0}^{1} x\left[\frac{N}{2}n_1 + \frac{N}{4}n_2 + n_3\right] \times W_N^{\left(\frac{N}{2}n_1 + \frac{N}{4}n_2 + n_3\right)(k_1 + 2k_2 + 4k_3)} \quad (9)$$

$$X[k_1 + 2k_2 + 4k_3] = \sum_{n_3=0}^{(N/4)-1} \sum_{n_2=0}^{1} \left[ B_{N/2}^{k_1}\left[\frac{N}{4}n_2 + n_3\right] W_n^{\left(\frac{N}{4}n_2 + n_3\right)} \right] \times W_N^{\left(\frac{N}{4}n_2 + n_3\right)(2k_2 + 4k_3)} \quad (10)$$

Where,

$$B_{N/2}^{k_1} = x\left[\frac{N}{4}n_2 + n_3\right] + (-1)^{k_1} x\left[\frac{N}{4}n_2 + n_3 + \frac{N}{2}\right] \quad (11)$$

For normal Radix-2 DIF FFT algorithm, the expression in the braces is computed first as a first stag in (10). However, in Radix $2^2$ FFT algorithm, the idea is to reconstruct the first stage and the second stage twiddle factors [16].

$$W_N^{\left(\frac{N}{4}n_2 + n_3\right)k_1} W_N^{\left(\frac{N}{4}n_2 + n_3\right)(2k_2 + 4k_3)}$$
$$= W_N^{Nn_2k_3} W_N^{Nn_2(k_1 + 2k_2)} W_N^{n_3(k_1 + 2k_2)} W_N^{4n_3k_3} \quad (12)$$
$$= (-j)^{n_2(k_1 + 2k_2)} W_N^{n_3(k_1 + 2k_2)} W_N^{4n_3k_3}$$

Observe that the last twiddle factor in (12) can be rewritten as:

$$W_N^{4n_3k_3} = e^{\frac{-j2\pi}{N}(4n_3k_3)} = e^{\frac{-j2\pi}{4N}(n_3k_3)} = W_{N/4}^{n_3k_3} \quad (13)$$

By applying (12), (13) and (10) and expand the summation over $n_2$, the result is a DFT definition with four times shorter FFT length.

$$X[k_1 + 2k_2 + 4k_3] = \sum_{n_3=0}^{(N/4)-1} \left[ H[k_1 + 2k_2 + 4k_3] W_n^{n_3(k_1 + 2k_2)} \right] W_N^{n_3k_3} \quad (14)$$

Where,

$$H[k_1 + 2k_2 + 4k_3] = \left[ x(n_3) + (-1)^{k_1} x\left(n_3 + \frac{N}{2}\right) \right] + (-j)^{(k_1 + 2k_2)} \left[ x\left(n_3 + \frac{N}{4}\right) + (-1)^{k_1} x\left(n_3 + \frac{3N}{4}\right) \right] \quad (15)$$

Equation 14 is known as Radix $2^2$ FFT algorithm. Fig. 2 shows the butterfly signal flow graph for radix $2^2$ FFT algorithms.

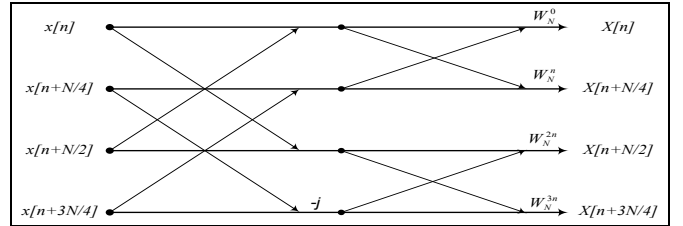

Fig. 2 butterfly signal flow graph for Radix $2^2$ FFT algorithm.

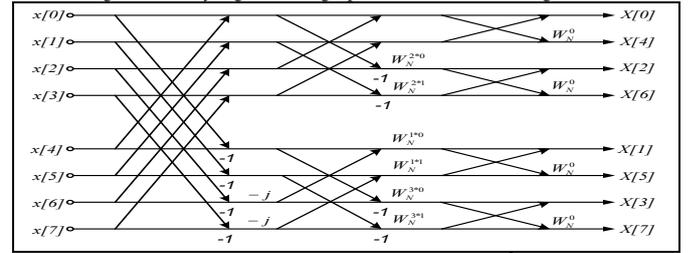

Fig. 3 Signal flow graph for an 8-Point Radix $2^2$ DIF FFT

Fig. 3 shows the Radix $2^2$ algorithm for an 8 point FFT. The complex multiplication in the first stage will be multiplication with (–j), which means just swapping the real with imaginary and sign inversion. One complex multiplier can be reduced for 8-point FFT implementation.

From (15), each stage in Radix $2^2$ SDF FFT consists of Butterfly I, Butterfly II, Complex multipliers with twiddle factors. Butterfly I calculate the input data flow, butterfly II calculate the output data flow from Butterfly I, than multiply the twiddle factors with the output data from Butterfly II, to get the result of the current stage. Fig. 4 shows the structure of 8 point Radix $2^2$ SDF FFT.

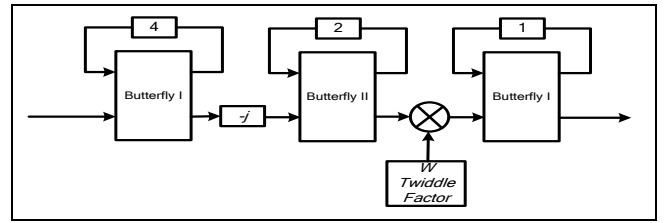

Fig. 4 8-Point Radix $2^2$ SDF DIF FFT Structure

*A. Butterfly I Structure*

Fig. 5 shows the Butterfly I structure, the input $A_r$, $A_i$ for this butterfly comes from the previous component which is the twiddle factor multiplier except the first stage it comes form the FFT input data. The output data $B_r$, $B_i$ goes to the next stage which is normally the Butterfly II. The control signal $C1$ has two options $C1=0$ to multiplexers direct the input data to the feedback registers until they filled. The other option is $C1=1$ the multiplexers select the output of the adders and subtracters.

*B. Butterfly II Structure*

Fig. 6 shows the Butterfly II structure. The input data $B_r$, $B_i$ comes from the previous component, Butterfly I. The output data from the Butterfly II are $E_r$, $E_i$, $F_r$ and $F_i$. $E_r$, $E_i$ fed to the next component, normally twiddle factor multiplier. The $F_r$ and $F_i$ go to the feedback registers.

The multiplication by –j involves swapping between real part and imaginary part and sign inversion. The swapping is handled by the multiplexers Swap-MUX efficiently and the sign inversion is handled by switching between the adding and the subtracting operations by mean of Swap-MUX. The control signals $C1$ and $C2$ will be one when there is a need for multiplication by −j, therefore the real and imaginary data will swap and the adding and subtracting operations will switched. In order to not lose any precision the divide by 2 is used where the word lengths imply successive growth as the data goes through adder, subtracter and multiplier operations. Rounding off has been also applied to reduce the scaling errors.

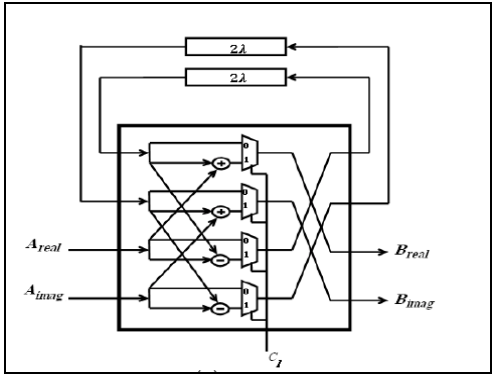

Fig. 5 Butterfly I

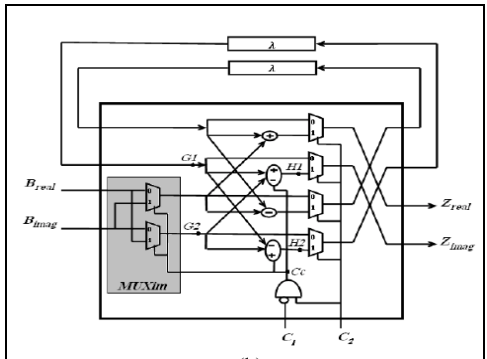

Fig. 6 Butterfly II

*C. Digit Slicing Complex Multiplier Less*

Complex multiplier can be realized by digit-slicing multiplier-less and real adder [17] based on (16) as shown in Fig. 7.

$(a_r+ja_i)(b_r+jb_i)=\{b_r[a_r-a_i]+a_i[b_r-b_i]\}+j\{b_i(a_r+a_i)+a_i(b_r-b_i)\}$  (16)

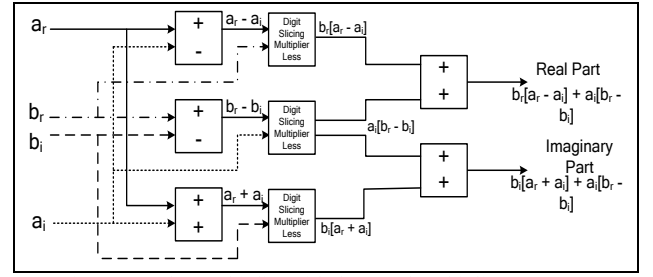

Fig. 7 Complex multiplier with three real multiplier structures

## IV. SHIFT AND ADD DIGIT SLICING MULTIPLIER LESS

The proposed design slicing the input data to four blocks each block carry four bits. by considering the input data for the multiplier are A and B with the word-length of 16 bits two's complement fixed point signed number with 15 bits fraction. The digit slicing architecture applied for the input A as shown in Fig 8. There are four different cases for the multiplication between the four bits and the twiddle factors. Fig. 8 shows the block diagram of the digit-slicing multiplier less using shift and addition technique.

Because of the shifts operation according to the digit slicing algorithm the twiddle factors will store with right shifts by 6 which means that the ROM for store the twiddle factors will be 10 bits width only not 16 bits. As mentioned in (4) and (5) the digit-slicing algorithm for this case will be:

$$A_3.B = \sum_{j=0}^{3} 2^j A_{3,j}.B \quad , \quad A_2.B = \sum_{j=0}^{3} 2^j A_{2,j}.B \quad ,$$

$$A_1.B = \sum_{j=0}^{3} 2^j A_{1,j}.B \quad , \quad A_0.B = \sum_{j=0}^{3} 2^j A_{0,j}.B$$

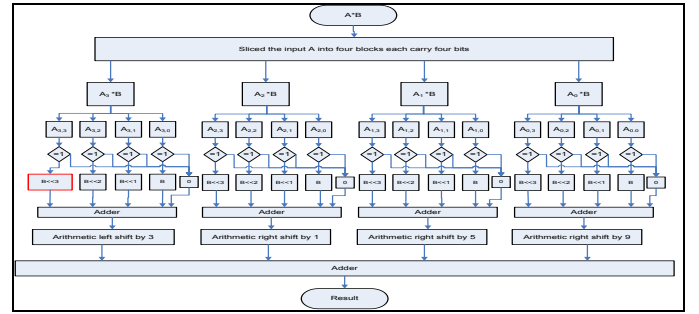

Fig. 8 The structure of Digit Slicing Multiplier Less with shift and add

## V. IMPLEMENTATION RESULT

The proposed design of pipeline Radix $2^2$ DIF SDF-FFT processor based on digit-slicing multiplier-less has been implemented using Matlab to prove and check the result for all stages as shown in Fig. 9. The design has been coded in Verilog HDL and tested in real hardware using Xilinx Virtex-4 FPGA as shown in Fig. 10 and Fig. 12.

In addition, the Modelsim XE-III was used to get the simulation result of the proposed design as shown in Fig. 11.

Table 1 shows the synthesis results with a comparison with conventional FFT processor similar work.

TABLE I
FFT COMPARISON

| Xilinx Virtex- 4 FPGA | Total gate count | Max.Freq. MHz |
|---|---|---|
| Conventional 8 point FFT [9] | 77,418 | 200 |
| Design Proposed by [9] | 16,580 | 400 |
| **Proposed Design** | **14,854** | **669.277** |

## VI. CONCLUSION

This study presented the FPGA Implementation of pipeline Radix $2^2$ DIF SDF-FFT processor based on digit-slicing multiplier-less. The implementation has been coded in Verilog HDL and was tested on Xilinx Virtex-4 FPGA prototyping board. A maximum clock frequency of 669.277 MHz with total equivalent gate count of 14,854 have been obtained from the synthesis report for the 8 point pipeline Radix $2^2$ DIF SDF FFT which is 3.35 time faster than the conventional butterfly and it required 20% of the conventional butterfly area. It can be concluded that the proposed design is an enabler in solving problems that affect communications capability in FFT and possesses huge potentials for future related research areas.

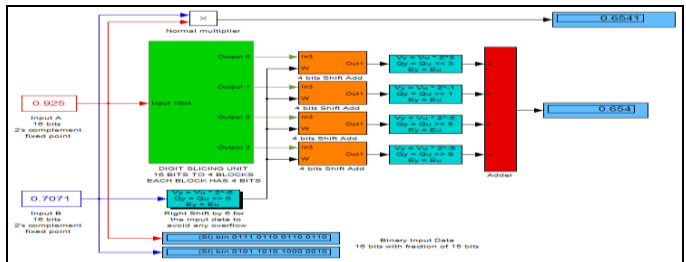

Fig. 9 Simulation of the 8-point pipeline digit slicing Radix $2^2$ SDF-FFT

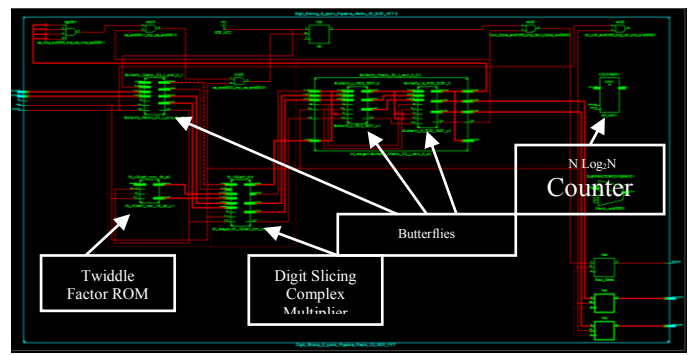

Fig. 10 Internal behavioral layout of top level Module of the 8-point pipeline digit slicing radix 22 SDF FFT

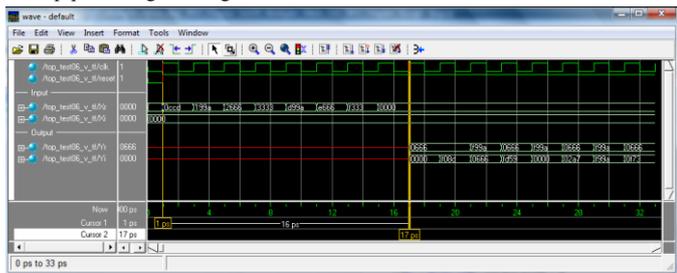

Fig. 11 Behavioral Simulation of the proposed design

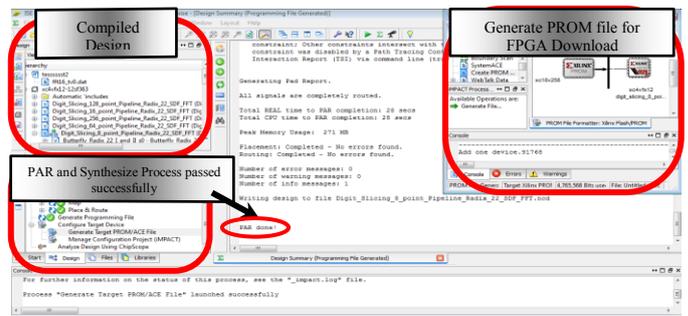

Fig. 12 FPGA synthesize report of the proposed design

REFERENCES

[1] A. V. Oppenheim, R. W. Schafer, and J. R. Buck, Discrete-time signal processing, 2 ed., N.J.: Prentice Hall, 1999.

[2] G. D. Bergland, "A radix-eight fast-Fourier transform subroutine for real-valued series.," IEEE Trans. Audio Electroacoust, vol. 17, pp. 138-144, 1969.

[3] R. C. Singleton, "An algorithm for computing the mixed radix fast Fourier transform," Audio and Electroacoustics, IEEE Transactions on vol. 17, pp. 93-103, 1969.

[4] D. P. Kolba and T. W. Parks, "A prime factor FFT algorithm using high-speed convolution," IEEE Trans Acoust. Speech, Signal Process, vol. 25, pp. 281-294, 1977.

[5] A. R. Varkonyi-Koczy, "A recursive Fast Fourier Transform algorithm," IEEE Trans.Circuits, vol. 42, pp. 614-616, 1995.

[6] Y.Wang,Y,Y.J.Tang,J.G.Chung,S.Song, "Novel memory reference reduction methods for FFT implementation on DSP processors," IEEE Trans. Signal Process, vol. 55, pp. 2338-2349, 2007.

[7] Y. Zhou,J.M. Noras, S. J. Shephend, "Novel design of multiplier-less FFT processors," Signal Proc., vol. 87, pp. 1402-1407, 2007.

[8] T. Sansaloni, A. P´erez-Pascual, V. Torres, and J. Valls, "Efficient pipeline FFT processors for WLAN MIMO-OFDM systems," Electronics Letters, vol. 41, pp. 1043–1044, 2005.

[9] B. Mahmud and M. Othman, "FPGA implementation of a canonical signed digit multiplier-less based FFT Processor for wireless communication applications," in ICSE2006 Proc Kuala Lumpur, Malaysia, 2006, pp. 641-645.

[10] B. M. Baas, "A Low-Power, High-Performance,1024-Point FFT Processor," IEEE JOURNAL OF SOLID-STATE CIRCUITS,, vol. 34, pp. 380-387, 1999.

[11] Y. P. Hsu and S. Y. Lin, "Parallel-computing approach for FFT implementation on Digital Signal Processor (DSP)," World Acad. Sci., Eng. Technol., vol. 42, pp. 587-591, 2008.

[12] M. A. B. Nun and M. E. Woodward, "A modular approach to the hardware implementation of digital filters " Radio and Electronic Engineer vol. 46, pp. 393 - 400 1976.

[13] A. Peled and B. Liu, Digital signal processing : theory, design, and implementation. New York: Wiley, 1976.

[14] Z. A. M. Sharrif, "Digit slicing architecture for real time digital filters." vol. Ph.D UK: Loughborough University, 1980.

[15] S. A. Samad, A. Ragoub, M. Othman, and Z. A. M. Shariff, "Implementation of a high speed Fast Fourier Transform VLSI chip " Microelectronics Journal, vol. 29, pp. 881-887 1998.

[16] S. He and M. Torkelson, "A new approach to pipeline FFT processor," in Parallel Processing Symposium, Proceedings of IPPS '96, The 10th International Honolulu, HI, 1996, pp. 766 - 770

[17] Rozita Teymourzadeh, Yazan Samir Algenabi, Nooshin Mahdavi, Masuri Bin Othman. On-Chip Implementation of High Resolution High Speed Floating Point Adder/Subtractor with Reducing Mean Latency for OFDM. American Journal of Engineering and Applied Sciences. 3(1): 25-30. ISSN: 1941-7020. 2010.

[18] Yazan Samir Algenabi,Rozita Teymourzadeh, Masuri Othman & Md Shabiul Islam, 2011. FPGA Implementation of pipeline Digit-Slicing Multiplier-Less Radix $2^2$ DIF SDF Butterfly for Fourier Transform Structure. *IEEE European Conference on Antennas and propagation*. pp 4168- 4172